*Journal of Risk and Financial Management*

MDPI

*Article*

# Econometric Modeling of Input-Driven Output Risk through a Versatile CES Production Function

Ali Zeytoon-Nejad [1,*] and Barry K. Goodwin [2]

[1] School of Business, Wake Forest University, Winston-Salem, NC 27109, USA; zeytoosa@wfu.edu
[2] Department of Economics and Department of Agricultural and Resource Economics, North Carolina State University; Raleigh, NC 27607, USA; barry_goodwin@ncsu.edu
\* Correspondence: zeytoosa@wfu.edu

**Abstract:** The conventional functional form of the Constant-Elasticity-of-Substitution (CES) production function is a general production function nesting a number of other forms of production functions. Examples of such functions include Leontief, Cobb–Douglas, and linear production functions. Nevertheless, the conventional form of the CES production specification is still restrictive in multiple aspects. One example is the fact that the marginal effect of increasing input use always has to be to increase the variability of output quantity by the conventional construction of this function. This paper proposes a generalized variant of the CES production function that allows for various input effects on the probability distribution of output. Failure to allow for this possible input–output risk structure is indeed one of the limitations of the conventional form of the CES production function. This limitation may result in false inferences about input-driven output risk. In light of this, the present paper proposes a solution to this problem. First, it is shown that the familiar CES formulation suffers from very restrictive structural assumptions regarding risk considerations, and that such restrictions may lead to biased and inefficient estimates of production quantity and production risk. Following the general theme of Just and Pope's approach, a CES-based production-function specification that overcomes this shortcoming of the original CES production function is introduced, and a three-stage Nonlinear Least-Squares (NLS) estimation procedure for the estimation of the proposed functional form is presented. To illustrate the proposed approaches in this paper, two empirical applications in irrigation and fertilizer response using the famous Hexem–Heady experimental dataset are provided. Finally, implications for modeling input-driven production risks are discussed.

**Citation:** Zeytoon-Nejad, Ali, and Barry K. Goodwin. 2023. Econometric Modeling of Input-Driven Output Risk through a Versatile CES Production Function. *Journal of Risk and Financial Management* 16: 100. https://doi.org/10.3390/jrfm16020100

Academic Editor(s): Thanasis Stengos

Received: 29 December 2022
Revised: 29 January 2023
Accepted: 1 February 2023
Published: date

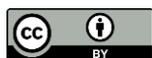

## 1. Introduction

Modeling production is a task of great importance in economics that may be conducted for many different purposes, such as the investigation of input-driven output risk. This important task is often carried out through the estimation of production functions. Production risk is an inseparable part of the production process in many economic sectors, including agriculture, in which context the empirical applications of this paper will be provided. As such, modeling output risk is an issue of great concern in the realm of economics. The Constant-Elasticity-of-Substitution (CES) production function is a popular specification in the estimation of production functions. The original, familiar specification of CES production function, as introduced by Arrow et al. (1961), is deemed to be a general specification that nests multiple types of production functions (i.e., Leontief, Cobb–Douglas, and linear). However, even this general specification of production functions is still restrictive in several aspects. This paper proposes a generalized variation of the CES production function that allows for various input effects on the probability distribution of output. Not allowing for this potential attribute can result in misleading conclusions and false econometric inferences about input-driven output risk. Therefore, the objective of the present paper is to investigate the essence of input–output response within the structure of the stochastic specification of the CES production function and generalize it in such a way that the resulting generalized CES production function can accommodate any type of input-driven output risk that may arise in real-world applications.

The generalization that the present paper proposes for the CES production function has to do with its inherent inflexibility with respect to input-driven output risks. It is common knowledge that production is a risky process that can involve various types of risks, including input-driven risks.[1] By definition, a production function is a mathematical relationship that relates the quantity of physical output of a production process to the quantities of its physical inputs. The deterministic specifications of production functions do not take into account the risky nature of production processes, since they consider production processes under certainty. In contrast, the stochastic specifications of production functions can effectively allow for the risky nature of production processes. Despite this, even the stochastic specifications of many commonly used production functions do not allow for flexibly modeling input effects on the probability distribution of output. Just and Pope (1979) have illustrated this inflexibility for the case of the Cobb–Douglas (1928) production function. However, the present paper attends to the CES production function, and it attempts to investigate and overcome a similar shortcoming for the case of the CES production function.

Arrow et al. (1961) introduced the CES production function for the first time and used it as a tool to investigate capital–labor substitution and economic efficiency. Uzawa (1962) used CES production functions to accommodate different elasticities of substitution in his specification, characterizing the class of production functions for which elasticities of substitution are all constant. Kmenta (1967) made one of the first and most serious efforts for the empirical estimation of the CES production function by providing estimation procedures applicable to the generalized version of the CES function, which allowed for the possibility of different non-constant returns to scale. Sato (1967) proposed the nested CES production function by devising a two-level CES production function. De La Grandville (1989) normalized the CES production function for a macroeconomic application to study economic growth, indicating that aggregate growth is faster and more sustainable once the elasticity of substitution between labor and capital is greater. León-Ledesma et al. (2010) normalized the CES production function and used a Monte Carlo analysis to characterize the conditions under which the identification of the CES function is viable and robust.

In addition to the usage of the CES specification as a production function, some scholars have also used the CES specification as a utility function, for example, Baumgärtner et al. (2013). Henningsen and Henningsen (2011) created an R package called micEconCES to facilitate the econometric estimation of the CES production and utility functions using the popular statistical software R. Zeytoon-Nejad et al. (2022) generalized the CES production function to allow for the inclusion of input thresholds within the structure of the CES production function. Jo and Miftakhova (2022) made use of the CES production function with an application in environmental economics, aiming to challenge the assumption of an exogenous, constant elasticity of substitution. Despite all the developments regarding the CES production function over its rather long lifetime and history, no one has ever attended to its risk implications, in particular its built-in construct of input-driven output risk, which is in fact the main purpose of the present paper.

Accordingly, the main objective of the present paper is to explore econometric possibilities for the CES production-function specification with typical risk implications. Indeed, this study considers the appropriate specification of the CES production function under risk and uncertainty. It is argued that the popular CES formulation of stochastic production functions is very restrictive for many cases in which modeling risk of production in the production function is of importance. In fact, when the CES production specification does not allow for various possible input effects on the probability distribution of output, the utilization of the CES production function imposes a strong structural assumption about input effects on the probability distribution of output (i.e., the marginal effect of increasing input use $\partial Var(y)/\partial x_i$ must always be to increase the variability of output), where "y" denotes output and "$x_i$" denotes input i. However, in reality, a reduction in the usage of some inputs (e.g., insecticides, herbicides, and irrigation in some cases) may cause more variable production.[2]

The remainder of this paper will proceed as follows. Section 2 is devoted to providing background on stochastic production functions in general and on the CES production function in particular and presenting the shortcoming of this production function with respect to its inflexibility regarding input effects on the distribution of output. Afterwards, a more generalized model specification is provided to overcome this deficiency, and the related estimation procedure is discussed. Section 3 introduces the empirical applications of the study, describes the data used in the analysis, and elaborates on the application of the methodology proposed in this paper. In addition, the results of the empirical applications and estimations are reported and discussed in this section. In Section 4, a conclusion is drawn, and plans for future research are discussed. Lastly, the paper ends with appendices to explain the derivations, procedures, estimations, and methods in greater detail.

## 2. Proposed Model and Estimation Method

Modeling production is an important undertaking in economics. It may be conducted for different purposes, including, but not limited to, addressing allocative efficiency in the usage of inputs in production, investigating the scale optimality of a production process, studying the productivity levels of various inputs over time, and evaluating and predicting the effects of government policies surrounding production-related regulations. This important task is often carried out through the estimation of production functions.

Production risk is an inseparable part of the production process in agriculture. As such, modeling output risk is an economic issue of great concern, because not only does the level of risk involved in a production process influence the optimal decisions to be made by economic agents, it may also determine the insurance premiums related to the production process as well as the interest expenses on the loans to be given to the production process of interest.

The formal specification of the CES production function is a popular specification in the estimation of production functions, but it is incapable of accommodating various types of input-driven output risk. In this section, we elaborate on this shortcoming of the CES production function and propose a generalized variant of the CES specification that overcomes this shortcoming.

The problem of examining stochastic facets of production response has been studied by some scholars in the past. For instance, Just and Pope (1979) have done so for the case of multiplicative production functions in general and for Cobb–Douglas and Translog production functions in particular. As another example, Asche and Tveterås (1999) have done so for the case of linear quadratic functional forms. Nonetheless, thus far, no one has paid adequate attention to the case of the CES production function. This is true of both theoretical and empirical studies.

As shown in this paper and particularly in this section, the CES production function, by construction, suffers from unrealistic structural assumptions concerning input-driven production risks. As a result of this inconsistency with some real-world phenomena, nearly all empirical and theoretical studies in which the CES production function is used as a tool to model the production process make implicit assumptions that increasing inputs usage always increases output risk. However, in reality, this is not necessarily true. It is evident that although increasing the quantity of some inputs (such as some fertilizers and land, *ceteris paribus*) can increase output risk, there are a number of other inputs (such as insecticides, herbicides, and irrigated acreage) whose increased quantities can reduce the level of output risk. To develop a general CES production function that allows for either of these possibilities, one can take advantage of the seminal estimation approach proposed by Just and Pope (1979). According to Just and Pope (1979), to attain such a generality, an adequate production-function specification should include two separate functions, one of which specifies the effects of input on the *mean* of output quantity, while the other specifies the effects of input use on the *variance* of output quantity. This section attends to this shortcoming of the CES production function and proposes a variation of the CES production function that overcomes the mentioned shortcoming.

In short, this section aims to examine the implications of the original econometric specification of the CES production function when input-driven output risk is of importance. In this section, a generalized variation of the CES production function for the estimation of its stochastic specification is proposed, and its generality in reflecting the risk effects of input use is illustrated. Finally, the respective estimation procedures are outlined. The functional form proposed in this paper is sufficiently general to embrace all the implications of the original CES production function but avoid the above-described shortcoming and unrealistic assumption that the original CES production-function specification imposes.

The traditional CES production function takes the following generic mathematical form:

$$y = A \left[ \sum_i \alpha_i x_i^r \right]^{\frac{1}{r}} e^\varepsilon \qquad (1)$$

where $y$ is output quantity, $x_i$ is the quantity of input $i$ and is always positive ($x_i > 0 \ \forall i$), $\alpha_i$ is the share parameter of input $i$ and is always positive ($\alpha_i > 0 \ \forall i$), $A$ is the level of total factor productivity or TFP ($A > 0$), $r$ is a function of the elasticity of substitution (i.e., $r = (s-1)/s$, and $s$ is the elasticity of substitution), and $\varepsilon$ is a stochastic disturbance, with $E(\varepsilon) = 0$ and $Var(\varepsilon) > 0$. Now, the variance of output quantity can be calculated as the following:

$$V(y) = A^2 \left[ \sum_i \alpha_i x_i^r \right]^{\frac{2}{r}} V(e^\varepsilon) \qquad (2)$$

Now, one can simply compute the marginal effect of input use on production variability as the following:

$$\frac{\partial V(y)}{\partial x_i} = 2 A^2 \alpha_i \left[ \sum_j \alpha_j x_j^r \right]^{\frac{2}{r}-1} x_i^{r-1} V(e^\varepsilon) > \mathbf{0} \qquad (3)$$

Therefore, due to the *structure* of the CES production function, the marginal effect of increasing input use always has to be to increase the variability of output quantity as long as the relevant $\alpha_i$ is positive. (Note that when marginal productivity is positive, $\alpha_i$ must be positive.) In practice, employing the commonly used formulation of the CES production function as reported in Equation (1) has some practical implications that are not in accordance with some real-world, economic phenomena. Following the example given by Just and Pope (1979), we consider the policy evaluation of limiting the usage of pesticides. Given the structure of the CES production-function specification introduced by Equation (1), a decrease in pesticide usage would imply a necessary reduction in output variability, according to Equation (3). However, in the real world, a reduction in the amount of pesticide used may result in more variable production (i.e., production with higher variance). Other examples of such an inconsistency between the *theory* produced by the CES production function and *practice* in the real world could include using large and fast harvesting equipment[3], applying frost and freeze protection tools (e.g., cloth covers, plastic covers, chemicals sprays, and smudge pots), and irrigated acreage.

Imposing this structural assumption, which is not necessarily true and overly restrictive, may cause some problems when it comes to making optimal decisions and formulating government policies. By definition, risk aversion is a preference for a sure (certain) outcome over a lottery with greater or equal expected value. Accordingly, a risk-averse agent is one that gains utility as a result of reducing outcome risk (here, "output" risk). However, in modeling production risk of inputs usage (such as pesticide) that reduces output risk, the original functional form of the CES production function incorrectly reports increased output risk, which in turn may result in making incorrect production-related decisions at an individual level as well as misleading conclusions on government policies about production-related regulations. To clarify matters and exemplify this issue, think of a risk-averse farmer. In this case, when the farmer increases the quantity of pesticide, the variability and risk of production are decreased. Since there is less risk involved now, the risk-averse farmer should gain some extra utility. However, the CES function, by construction, incorrectly reports an increase in the variability and risk of production (as Equation (3) implies), which will be translated as a utility loss associated with the incorrectly reported higher risk.

Additionally, the marginal effect of input quantity on the variability of marginal productivity can be investigated both in theory (derived from the CES production function) and in application (in real-world practices). To this end, the marginal productivity of input $i$ within the context of the CES production function can be calculated as the following:

$$\frac{\partial y}{\partial x_i} = A \alpha_i \left[ \sum_j \alpha_j x_j^r \right]^{\frac{1}{r}-1} x_i^{r-1} e^{\varepsilon} \qquad (4)$$

Furthermore, the variance of the marginal productivity of input $i$ can be computed as follows:

$$V\left(\frac{\partial y}{\partial x_i}\right) = A^2 \alpha_i^2 \left[ \sum_j \alpha_j x_j^r \right]^{\frac{2}{r}-2} x_i^{2(r-1)} V(e^{\varepsilon}) \qquad (5)$$

Now, one can simply obtain the marginal effect of input quantity on the variability of marginal productivity as the following:

$$\partial V\left(\frac{\partial y}{\partial x_i}\right) / \partial x_i = 2 A^2 \alpha_i^2 \left[ \sum_j \alpha_j x_j^r \right]^{\frac{2}{r}-3} x_i^{2r-3} (r-1) \left[ \sum_{j \neq i} \alpha_i x_i^r \right] V(e^{\varepsilon}) < 0 \qquad (6)$$

If we assume that the CES production function specified in Equation (1) exhibits diminishing marginal product with respect to input $i$ (which is a reasonable, conventional assumption, implying the concavity of the production function, or in fact the concavity of $E(y)$ in $x_i$), then the marginal effect of increasing input quantity is to necessarily decrease the variability of marginal productivity, which is not necessarily true in some real-world practices. In many realistic applications, the marginal effect of increasing input quantity could be to increase the variability of marginal productivity. A typical example of inputs of this sort is land, as explained by Just and Pope (1979). It is known that when a farmer increases the land acreage, ceteris paribus, the variability of the marginal productivity of land increases. According to Radner and Rothschild (1975), when a farmer operates on a larger land area, holding the other inputs constant, his or her production is more subject to adverse weather conditions during vital phases of production such as harvesting (during which crop is highly sensitive) and planting (during which seeds are highly sensitive). In brief, the *decreasing* marginal effect of increasing input quantity on the variability of marginal productivity is an overly restrictive structural assumption inherent in the original specification of the CES production function. Thus, it should be generalized somehow in order that it can accommodate many real-world situations better and more effectively.

To achieve the two important goals mentioned above, a more general stochastic specification should be proposed that is free of the two a priori restrictions. As Just and Pope (1979) argue, "the effects of input on output should not be tied to the effects of input on variability of output a priori". Following their seminal approach to overcoming such specification issues, one can work with a sensible production-function specification that include two separate functions, one of which specifies the effects of input on the *mean* of output quantity, while the other specifies the effects of input use on the *variance* of output quantity. As they have shown, such a production-function specification takes the following generic form:

$$y = f(X) + h^{\frac{1}{2}}(X)\varepsilon, \quad E(\varepsilon) = 0, V(\varepsilon) = 1 \quad (7)$$

Therefore, after this transformation, $E(y) = f(X)$ and $V(y) = h(X)$.[4] As a result, the effects of input on output and marginal output are no longer tied to the effects of input on the variability of output and marginal output *a priori*.[5]

Following Arrow et al. (1961) and Just and Pope (1979), a generalized version of the classical CES function with two inputs that allows for different types of input-driven output risks would take the following functional form:

$$y = A[\alpha_1 x^{r_1} + \alpha_2 z^{r_1}]^{\frac{1}{r_1}} + B[\beta_1 x^{r_2} + \beta_2 z^{r_2}]^{\frac{1}{2r_2}} \varepsilon \quad (8)$$

In what follows, an econometric approach will be put forth that outlines how to deal with the shortcoming characterized in Section 2. This generalization can also be applied and extended to other generalizations of the CES production function.[6]

The CES production function is inherently nonlinear in parameters. As such, it cannot be linearized analytically, and thus, it is impossible to estimate its formal specification using the common linear estimation methods. Therefore, the CES production function is often estimated by Nonlinear Least-Squares (NLS) estimation method, or alternatively, it is approximated by certain approximation methods, such as the so-called "Kmenta approximation", as put forth by Kmenta (1967), or "Uebe approximation", as proposed by Uebe (2000), which is still a second-order Taylor series expansion but slightly different from that of Kmenta, both of which can be estimated by linear estimation methods.[7] After all, since our focus in this paper is on the shortcomings of the *original* CES production-function specification, and also because the most straightforward way to estimate the CES production function is often the NLS method using different optimization algorithms,[8] we do not use the approximation methods with linear estimation approaches, and instead we work directly with the nonlinear CES production-function specifications and estimate them using the NLS method. The functional form of the CES that is usually estimated by the NLS method is as follows:

$$\ln(y) = \ln(A) + \frac{1}{r_1} \ln\left[\alpha_1(x)^{r_1} + \alpha_2(z)^{r_1}\right] + \varepsilon \quad (9)$$

As for the generalized variation of the CES production function that allows for input thresholds, one can likewise use the NLS method for the estimation of the following functional form:

$$\ln(y) = \ln(A) + \frac{1}{r_1} \ln\left[\alpha_1(x - b_1)^{r_1} + \alpha_2(z - b_2)^{r_1}\right] + \varepsilon \quad (10)$$

As for the generalized variation of the CES production function that allows for the flexibility of input-driven output risk, we follow Just and Pope (1979)'s approach. If one considers Equation (7) and those reported in Appendix A, Just and Pope (1979)'s approach can be summarized through the following three-step estimation procedure:

(1) An NLS regression of $y_t$ on $f(X_t, \alpha)$, obtaining $\hat{\alpha}$.
(2) An NLS regression of $(\hat{\varepsilon}^*)^2 = (y_t - f(X_t, \hat{\alpha}))^2$ on $h(X_t, \beta)$, obtaining, $\hat{\beta}$.
(3) An NLS regression of $y_t^* = y_t h^{-1/2}(X_t, \hat{\beta})$ on $f^*(X_t, \alpha) = f(X_t, \alpha) h^{-1/2}(X_t, \hat{\beta})$, obtaining $\hat{\alpha}$.

[Appendix B provides greater details on the specifics of this three-step estimation approach.]

## 3. Empirical Results and Discussion

Although production functions such as the CES are per se "technical" relationships between input quantities and output quantities, adding the above-mentioned real-world economic component while retaining its conventional desired properties will enrich the economic implications of this production function. This paper aims to enhance the economics of the CES production function by adding this supplementary risk aspect to the standard form of the CES production function, thereby allowing for the flexibility to account for different types of input-driven output risks. We allow for this generalization in empirical applications using several field trials from the famous Hexem–Heady (1978) dataset that contains experimental data on irrigation and fertilizer response.

This section presents two empirical applications in order to provide empirical support for the generalization introduced in this paper (i.e., the flexibility of input-driven output risk) for generalizing the classical CES production function. The well-known Hexem–Heady experimental dataset is used for our empirical purposes. The Hexem–Heady yield dataset has been created for several crops through controlled experiments by varying nitrogen and irrigation over multiple fixed levels and over a few years. These data are useful for exemplifying the generalization proposed in this paper for three reasons: First of all, there will not be any problems of multi-collinearity in this dataset. Secondly, there will not be any problems related to endogeneity in this dataset, as the experimental design is orthogonal. Third, it will be feasible to concentrate on each input separately when investigating their risk effects, as it is a controlled experiment. The dataset consists of a cross-section of several experimental plots over a few years. This dataset was generated with a cross-section of experimental plots. As the plots of each crop/location were located in the same site, the plots are very close to each other, so we do not need to specify any dummy variables for plot. Additionally, the plot attributes were not recorded during the course of data collection, so it is not even possible to specify plot dummies. However, as plots are impacted by the same weather-related conditions each year, time effects are very likely to be of importance in this dataset. Therefore, we need to include year dummies to pick up time effects and differences across years. We use Hexem–Heady experimental datasets for wheat at Yuma Mesa, AZ, as well as corn in Colby, KS. Table 1 provides the descriptive statistics of the variables existing in the above-mentioned datasets.

**Table 1.** Descriptive Statistics of Variables Used for Estimating CES Production Functions by Just and Pope's Approach.

| Experiment Station/Crop | Year(s) | Variables | Number of Observations | Sample Summary Statistics | | | |
|---|---|---|---|---|---|---|---|
| | | | | Mean | Standard Deviation | Minimum | Maximum |
| Yuma Mesa, AZ /Wheat | 1970–1971 | Water | 88 | 25.23 | 8.66 | 12.0 | 42.4 |
| | | Nitrogen | 88 | 155.68 | 108.26 | 0.0 | 325.0 |
| | | Yield | 88 | 2207.89 | 1449.66 | 479.0 | 6050.0 |
| Colby, KS /Corn | 1970–1971 | Water | 88 | 15.12 | 5.47 | 8.2 | 27.6 |
| | | Nitrogen | 88 | 180.00 | 139.16 | 0.0 | 360.0 |
| | | Yield | 88 | 6391.16 | 2292.84 | 956.0 | 10409.0 |

Note: The measurement units of the variables of interest are as follows water (acre-inches), nitrogen (pounds per acre), wheat yield (pounds of wheat per acre), and corn yield (pounds per acre of corn grain). Due to the fact that the CES production-function specification converges to a multiplicative functional form as the parameter r approaches zero ($r \to 0$), it makes sense to consider nonzero values for input quantities. Hence, 1 was added to all of the numerical values of nitrogen in the estimations, in order to avoid potential technical issues when the CES production-function specification converges to a multiplicative form.

In order to estimate the generalized variant of the CES production function that makes full allowance for different types of input-driven output risks, as explained in Section 2, following Arrow et al. (1961) as well as Just and Pope (1979), and considering the specific features of the dataset of interest, we can estimate the following functional form:

$$y_{ti} = A[\alpha_1 x_{ti}^{r_1} + \alpha_2 z_{ti}^{r_1}]^{\frac{1}{r_1}} + B[\beta_1 x_{ti}^{r_2} + \beta_2 z_{ti}^{r_2}]^{\frac{1}{r_2}} (\epsilon_{ti} + \omega_t) \quad (11)$$

Just and Pope (1979) have introduced and taken this approach for the first time to estimate Cobb–Douglas and Translog production functions, while we have used their approach to estimate CES production functions, which is considered to be a more general production-function specification. Nobody has thus far applied Just and Pope's approach to the CES production function, which per se is considered to be a general production function, nesting Cobb–Douglas, Leontief, and linear functions as its special cases. Estimating such a general production function through Just and Pope's approach will enrich the CES production specification by generalizing it one step further, in the sense that the resulting generalized variant of the CES will account for the flexibility of input-driven output risk, the absence of which is a shortcoming of the classical CES production-function specification. Additionally, Just and Pope (1979) consider only one input (i.e., fertilizer) in their production-function specifications, while we consider two inputs (i.e., irrigation water and nitrogen), each of which could potentially have a different risk implication. We have done so because we believe that it would be more interesting to see the workings and applicability of their general approach when more than one input is involved in modeling the production process of interest.

Equation (11) is, in fact, an equivalent version of Equations (7) and (8) for the CES production-function specification, although it has been modified to account for the fact that the data are generated experimentally by a cross-section of time series, where potentially the same random phenomena (e.g., weather-related variables) have affected all the contemporaneous observations. Under such circumstances, time effects can alternatively be captured through the use of time-dummy variables that represent different time periods.[9] For this reason, our model is augmented with a time-dummy variable. Thereby, we no longer have to deal with the *t* subscripts, since we have added the time aspect of the data to the model as a year dummy. Therefore, we end up with the following equation:

$$y_i = Ae^{A_{1971}D_{1971}}[\alpha_1 x_i^{r_1} + \alpha_2 z_i^{r_1}]^{\frac{1}{r_1}} + Be^{B_{1971}D_{1971}}[\beta_1 x_i^{r_2} + \beta_2 z_i^{r_2}]^{\frac{1}{r_2}} \epsilon_i \qquad (12)$$

By using logarithms in a stage-by-stage fashion[10], the above equation turns into the following logged functional form:

$$\ln(y_i) = \{\ln(A) + A_{1971}D_{1971} + \frac{1}{r_1}\ln[\alpha_1 x_i^{r_1} + \alpha_2 z_i^{r_1}]\} + \{\ln(B) + B_{1971}D_{1971} + \frac{1}{r_2}\ln[\beta_1 x_i^{r_2} + \beta_2 z_i^{r_2}] + \varepsilon_i\} \qquad (13)$$

Henningsen and Henningsen (2011) emphasize that a CES production function must be consistent with economic theory, so it is desired to assume that the summation of the factor shares is equal to unity. We make this assumption for its theoretical support as well as to control the potential exotic behavior of the CES production function due to its non-linear nature and the parameter *r* that exists within the exponent of the function. We do not impose any other restrictive assumptions in our estimations of the CES production functions in this section, mainly because a primary aim of the present paper is to *generalize* the essentially general production function, i.e., the CES production function.

Following the same estimation procedure as outlined in Section 2 and elaborated further in Appendix B, we estimated our generalized CES production-function specification. Table 2 reports the NLS estimates pertaining to the first stage of the estimation procedure of the generalized CES function for wheat at Yuma Mesa, AZ, as well as corn in Colby, KS.

**Table 2.** First-Stage Estimates of the Deterministic Component of Production Modeled by the CES Specification.

| Crop | Station | | Constant Term | Year Dummy | Coefficients | | |
|------|---------|---|---|---|---|---|---|
| | | | ln(A) | $A_{1971}$ | $r_1$ | $alpha_1$ | $alpha_2$ |
| Wheat | Yuma Mesa, AZ | Estimate | 3.4554 *** | 0.8805 *** | 0.4094 | 0.7840 *** | 0.2160 *** |
| | | Standard Error | (0.0872) | (0.1007) | (0.2560) | (0.0339) | (0.0339) |
| Corn | Colby, KS | Estimate | 5.7330 *** | −0.1577 * | 0.4441 * | 0.8855 *** | 0.1145 *** |
| | | Standard Error | (0.0760) | (0.0683) | (0.1932) | (0.0249) | (0.0249) |

Asterisks (*, **, and ***) show statistical significance at the 5%, 1%, and 0.1% levels, respectively (i.e., * $p < 0.05$, ** $p < 0.01$, *** $p < 0.001$).

As reported in Table 2, all the estimates are statistically significant and economically meaningful (even the estimate of *$r_1$* for wheat is statistically significant, at a significance level close to 10%). For both crops, the *r* parameters are somewhat close to zero, implying that the elasticity of substitution falls in the neighborhood of 1, indicating that the production function is fairly close in form to the case of Cobb–Douglas. More precisely, the shapes of the estimated CES production functions fall between the Cobb–Douglas and linear specifications (with $0 < r < 1$ or $1 < s < +\infty$), implying that water and nitrogen are relative substitutes in both cases. Regarding the factor-share parameters, for both crops, the share of water is considerably larger than that of nitrogen.

This is indeed the endpoint of the first stage of the estimation procedure. As explained in Appendix B, when the main objective of the estimation is to gain a general understanding of solely $f(X_t, \alpha)$ (i.e., the "mean" of output), the end of the first stage can be viewed as the endpoint of the estimation and analysis. Nonetheless, there are several potential rationales for proceeding to the second and third stages as well. As outlined by Just and Pope (1979), such reasons can include the following: (1) learning more about the marginal effect of inputs use on output risk, (2) performing more reliable hypothesis testing if there is a possibility of heteroscedasticity, and/or (3) gaining more efficiency in estimation, at least asymptotically. As one of the primary goals of the present paper is to explore the marginal effects of inputs use on output risk, we certainly have to proceed to the second and third stages of the estimation procedure as well; however, we examine the possibility of heteroscedasticity in the dataset to possibly find additional reasons to proceed to the second and third stages. As reported in Appendix C and shown in Figures A1–A4, our investigation on heteroscedasticity indicates that (1) there exists a sizable degree of heteroscedasticity in the data, implying that any statistical inference based on the results of the related regression model could be imprecise, and more importantly, (2) the type of the existing heteroscedasticity in most cases is *decreasing* in inputs, which cannot be addressed and accounted for accurately when one uses the original CES production-function specification, as shown and proved in Section 2 by Equation (3). To resolve these issues, we need to take the analysis to the second and third stages in order to develop a precise understanding of the true effects of inputs use on output mean and variance, as proposed by Just and Pope (1979).[11]

Table 3 contains the NLS estimates of the second-stage estimation of Just and Pope's procedure, which investigates the effect of inputs use on the *variability* (i.e., variance) of output for the case of wheat production at Yuma Mesa, AZ, as well as corn production in Colby, KS. As such, the results obtained from the second-stage estimation are not associated with a regular, classical production function, which is typically concerned mostly with the mean of output. In other words, the results reported in Table 3 are associated with a production *variability* function, and therefore, the usual economic-theoretical conditions that normally must hold true for any production (mean) function (e.g., positive factor shares) do not have to necessarily hold true for the production variability function estimated in the second stage and reported in Table 3.[12]

**Table 3.** Second-Stage Estimates of the Stochastic Component of Production Modeled by the CES Specification.

| Crop | Station | | Constant Term | Year Dummy | | Coefficients | |
|---|---|---|---|---|---|---|---|
| | | | ln(B) | $B_{1971}$ | $r_2$ | $beta_1$ | $beta_2$ |
| Wheat | Yuma Mesa, AZ | Estimate | −6.0578 *** | 0.9224 *** | −0.2797 | 1.0430 *** | −0.0430 *** |
| | | Standard Error | (0.0870) | (0.0470) | (0.3755) | (0.0122) | (0.0122) |
| Corn | Colby, KS | Estimate | −4.4035 *** | 0.3236 | −0.2535 | 1.0219 *** | −0.0219 * |
| | | Standard Error | (0.3805) | (0.1664) | (2.3711) | (0.0110) | (0.0110) |

Asterisks (*, **, and ***) show statistical significance at the 5%, 1%, and 0.1% levels, respectively (i.e., * p < 0.05, ** p < 0.01, *** p < 0.001).

As mentioned earlier, the results reported in Table 3 are related to production variability functions and not related to production (mean) functions. In other words, the classical, theoretical conditions desired for production functions must hold for an estimation specification that represents the *mean* of a production process, while the functions estimated at the second stage represent the *variance* of a production process. Therefore, the fact that the signs of the estimated betas are negative or the magnitudes of betas are greater than one in some cases is not troubling at all in the second stage. The estimated beta coefficient ($\beta_i$) can be interpreted as the marginal effect of input $i$ on output "variability". As such, unlike an estimated factor share ($\alpha_i$), which is always assumed to be positive ($\alpha_i > 0\ \forall i$) in the relevant range, estimated beta coefficients can meaningfully be positive or negative. As shown in Table 3, the marginal effect of water on yield variability for the cases of both wheat and corn in the related regions is statistically significantly positive, while the marginal effect of nitrogen use on yield variability for the cases of both wheat and corn has been estimated to be trivially negative and very close to zero.[13] Our results about nitrogen's effect are in line with that of Just and Pope (1979)'s empirical application of oats using the well-known Day dataset, in which the marginal effect of fertilizer use on oats yield variability turned out to be negative in their estimated Translog production function. The findings reported in Table 3 can be interpreted as follows. In the datasets of interest, increasing water irrigation causes the variance of output to increase, while increasing nitrogen use, on average, causes almost no change in the variance of output. When these results are compared with the *actual* data (through a scatterplot that depicts the relationship between the data on inputs and yield), it becomes apparent that these findings are intuitive and accurate. Figure 1 provides such visual checks.[14]

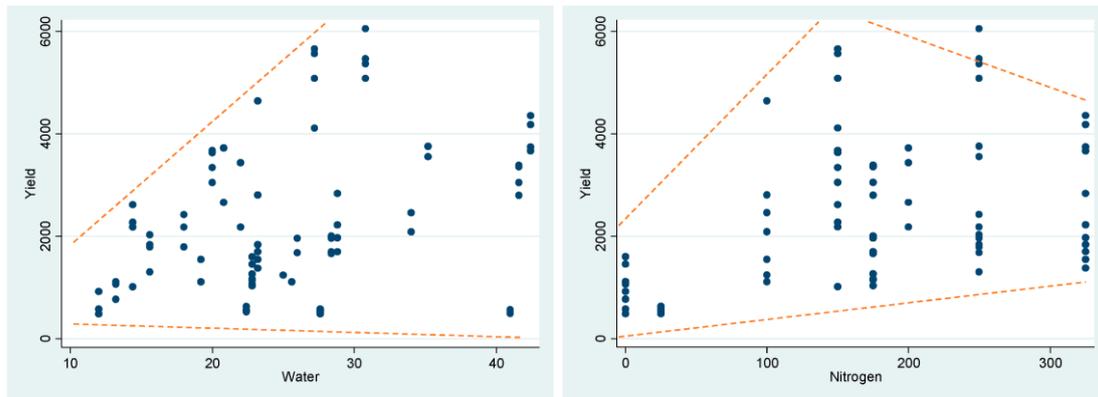

**Figure 1.** Scatterplot of Yield versus Input Values for Wheat at Yuma Mesa, AZ.

As clearly depicted by Figure 1, for the case of wheat at Yuma Mesa, AZ, the variance of wheat yield is greater for greater values of water than that for smaller values of water, meaning that increasing irrigation water use causes the variance of output to increase.[15] However, roughly speaking, the variance of yield is larger for intermediate values of nitrogen than that for extreme values of it. These two offsetting forces imply that increasing nitrogen use, on average, causes no significant change in the variance of wheat yield at Yuma Mesa, AZ. In other words, the two scatterplots above indicate that water has an increasing effect on the variability of output (i.e., wheat yield) and nitrogen has almost no effect on the variability of wheat output, both of which are exactly consistent with beta estimates reported in Table 3.[16] Figure 2 provides similar visual checks but for the case of corn in Colby, KS.

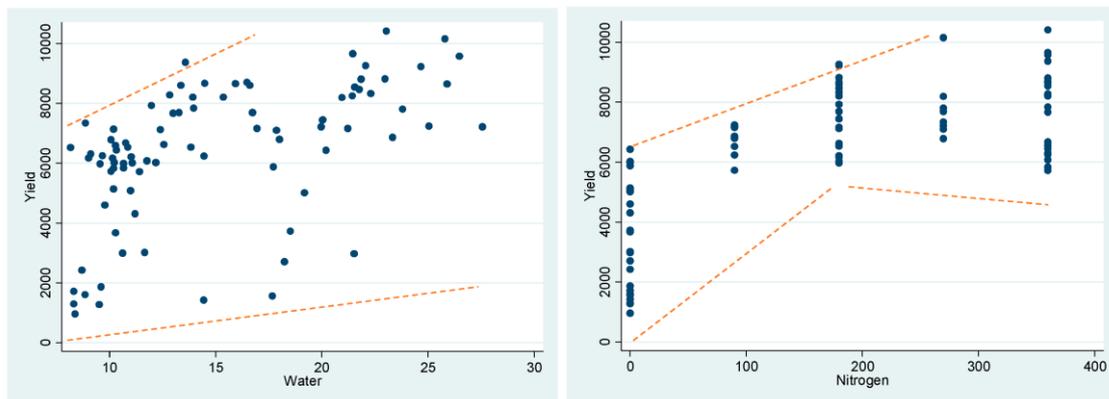

**Figure 2.** Scatterplot of Yield versus Input Values for Corn in Colby, KS.

As evidently showed by Figure 2, for the case of corn in Colby, KS, the variance of corn output quantity is larger for larger values of water than that for smaller values of water, implying that increasing irrigation water use causes the variance of output to increase.[17] However, roughly speaking, the variance of yield is smaller for intermediate values of nitrogen use than that for very small and very large values of it, causing an *hourglass-shaped* heteroscedasticity form. These two opposite effects imply that increasing nitrogen use, on average, causes no significant change in the variance of corn yield in Colby, KS.[18] In sum, the two scatterplots above indicate that water has an increasing effect on the variability of corn yield and nitrogen has a trivial, negative effect on the variability of corn yield, both of which are exactly consistent with beta estimates reported in Table 3.

As explained by the sign determination of Equation (6), the traditional CES, by construction, is not capable of addressing *decreasing* heteroscedasticity, which is true of nitrogen for both of the empirical applications of the present paper, although they are slightly negative. As a result, we need to disjoin the effect of the marginal use of input on the mean of output from that on the variability of output, as suggested by Just and Pope (1979) and explained previously. The next step to achieve this goal is to estimate the third stage of Just and Pope's approach. Table 4 reports the results of the third-stage estimation.

**Table 4.** Third-Stage Estimates of the Deterministic Component of Production Modeled by the CES Specification.

| Crop | Station | | Constant Term | Year Dummy | Coefficients | | |
|---|---|---|---|---|---|---|---|
| | | | Ln(A) | $A_{1971}$ | $r_1$ | $alpha_1$ | $alpha_2$ |
| Wheat | Yuma Mesa, AZ | Estimate | 3.0302 *** | 0.6231 *** | −5.5546 | 0.0011 | 0.9989 *** |
| | | Standard Error | (0.0656) | (0.0696) | (4.6459) | (0.0060) | (0.0060) |
| Corn | Colby, KS | Estimate | 4.5437 *** | 0.1516 ** | −1.3264 * | 0.0876 | 0.9124 *** |
| | | Standard Error | (0.3129) | (0.0493) | (0.6021) | (0.1216) | (0.1216) |

Asterisks (*, **, and ***) show statistical significance at the 5%, 1%, and 0.1% levels, respectively (i.e., * $p < 0.05$, ** $p < 0.01$, *** $p < 0.001$).

The third-stage estimates have been made such that they account for (1) the heteroscedasticity existing in the dataset as well as (2) the inability of the CES production-function specification to account for decreasing input-driven output risk. As a result, the estimates are somewhat different from those obtained in the first stage. The differences can be justified by understanding the fact that the factor-share estimates of the first stage are obtained in such a way that they contain information about both the mean and variance of output (i.e., as proxies for output means, they are confounded by the effect of output variance as well), while the factor-share estimates of the third stage are obtained under the condition that they contain information only about the net effect of inputs use on output means and not output variance anymore (i.e., as proxies for output means, they are no longer confounded by the effect of output variance). This is because the output variance was separately attended to and addressed in the second stage and before arriving at the third stage, so that effect was captured in the second-stage estimation and was not carried over into the third stage. Therefore, it does not make much sense to compare the results of the first stage and the third stage, and even if one intends to do so, they should compare the results of the first stage with those from the combined results of the second and third stage, which corresponds to the production-function specification introduced by Equations (8) and (12).

Furthermore, it is interesting to compare the standard-error estimates reported in Tables 2 and 4 (even though they only apply asymptotically). It is readily found that, except for constant terms, standard-error estimates are, in most cases, greater in Table 4 than in Table 2. However, the third-stage estimator is asymptotically efficient and thus possesses lower standard errors (asymptotically) than the first-stage estimator, as elaborated by Just and Pope (1979). That is to say, the *true* standard-error estimates reported in Table 2 (of the first stage) should be larger than those reported in Table 4. In fact, the standard-error estimates listed in Table 2 are not applicable under the general stochastic specification used in Table 4 estimates. In spite of this fact, unfortunately, in econometric practice, it is often the case that econometricians unrealistically assume homoscedasticity when risk facets of a study are presumed unimportant or information to the contrary is not readily apparent, as discussed by Just and Pope (1979). The comparison of Tables 2 and 4 demonstrates and exemplifies the possible hazard of assuming homoscedasticity when such an assumption does not hold true in reality. That is, when Just and Pope's approach to modeling production (i.e., Equation (7)) is taken, the conventional practice of assuming homoscedasticity (which may or may not align with reality) could correspond to the estimates resulting from the first stage of Just and Pope's estimation procedure. However, the resulting statistics from the first stage would be misleading and suggest far more precision in estimation than what is indeed warranted, as illustrated through the comparison of Tables 2 and 4. Therefore, employing the original CES production-function specification in modeling production risk can incorrectly report input-driven output risks, which in turn may result in making incorrect production-related decisions at an individual level as well as misleading conclusions about government production-related policies and regulations.

## 4. Summary and Conclusions

The CES production function is one of the most popular production functions in economics. This is primarily because of the high level of flexibility and generality that it provides compared to those of other production functions. However, the CES production function is still restrictive in that its construction allows only for modeling production contexts in which increasing inputs usage increases output risk. This paper generalizes the CES production function to allow for the flexibility of input-driven output risk. Accordingly, the objective of the paper is to investigate input–output response within the structure of the stochastic specification of the CES production function and generalize it in such a way that it accommodates various types of input-driven output risk. Not allowing for catering to various types of input-driven output risk is indeed a shortcoming of the classical CES production-function specification, which in turn could lead to invalid conclusions and false inferences about the essence of input-driven output risk. The paper starts with providing theoretical reasoning and motivations for the generalized variation of the CES production-function specification to be introduced, and afterwards, it continues with providing empirical applications and support for the generalization proposed.

It is shown that the familiar CES formulation suffers from very restrictive structural assumptions regarding risk considerations. As a result, nearly all empirical and theoretical studies in which the CES production function is used as an instrument to model production processes make implicit assumptions that increasing inputs usage always increases output risk. However, in reality, this is not necessarily true. It is further shown that such restrictions may lead to biased and inefficient estimates of production quantity (i.e., mean) and production risk (i.e., variance). Following Just and Pope (1979), a production-function specification that overcomes this shortcoming of the CES production function (in which the effects of input on output are no longer tied to the effects of input on variability of output a priori) is introduced, and a three-stage Nonlinear Least-Squares (NLS) estimation procedure for the estimation of the proposed functional form is presented. To illustrate the practicability of this generalization in real-world applications, two empirical applications are provided. Just and Pope (1979) have illustrated this inflexibility for the case of Cobb–Douglas production function and have provided empirical applications for production functions with only one single input. However, the present paper attends to the CES production function, attempts to investigate and overcome the same shortcoming for the case of the CES production function, and additionally, considers two inputs (i.e., irrigation water and nitrogen), each of which could potentially have a different risk implication, in order to better see the workings and applicability of the approach when more than one input is involved.

As shown in this paper, the estimates from this generalized approach to estimating the CES specification have been made such that they account for (1) the heteroscedasticity existing in the dataset as well as (2) the inability of the CES production-function specification to account for decreasing input-driven output risk. As a result, the estimates are somewhat different from those obtained from the estimation of the original CES specification, mainly due to the fact that the factor-share estimates of the first stage are obtained in such a way that they contain information about both the mean and variance of output (i.e., as proxies for output means, they are confounded by the effect of output variance as well), while the factor-share estimates of the third stage are obtained under the condition that they contain information only about the net effect of inputs use on output means and not output variance anymore (i.e., as proxies for output means, they are no longer confounded by the effect of output variance). This is because the output variance was separately attended to and addressed in the second stage and before arriving at the third stage, so that effect was captured in the second-stage estimation and was not carried over into the third stage.

Furthermore, the standard-error estimates are compared, leading to the conclusion that except for constant terms, standard-error estimates are, in most cases, greater than those from the original CES specification. However, the third-stage estimator is asymptotically efficient and thus possesses lower standard errors (asymptotically) than the first-stage estimator, as elaborated by Just and Pope (1979). That is to say, the *true* standard-error estimates of the original CES function should be larger than those estimated through our generalized CES specification. Therefore, the standard-error estimates resulting from the original CES specification are not applicable under the general stochastic specification of the CES production function. Therefore, the resulting statistics from the estimation of the original CES production-function specification may be misleading, suggesting much more precision in estimation than what is indeed warranted.

To sum up, we suggest that production modelers who intend to estimate the CES production-function specification should estimate the generalized variation of the CES specification proposed in this paper when risk considerations are important and heteroscedasticity is troublesome. If neither of the above-mentioned problems is a cause for concern in the setting of their analyses, then they should use the original CES specification. Additionally, Moosavian (2019) and Zeytoon-Nejad et al. (2022) have provided a generalized variant of the CES production function that accommodates input thresholds within the structure of the CES production function. In future research, modelers can employ the joint estimation of both the generalization introduced in this paper and that introduced by Moosavian (2019) and further discussed by Zeytoon-Nejad et al. (2022) if both of the aspects mentioned above are of great concern at the same time. Another possible reason for doing so could be that input thresholds and input-driven output risk can justifiably be thought of as two interconnected and interdependent aspects, and thus, one may want to reasonably investigate the two mentioned generalizations simultaneously and jointly. Zeytoon-Nejad et al. (2022) provide an appendix detailing the technical aspects of such a joint estimation. Another interesting area for future research can be applying the idea of a nested CES model to the model proposed in this paper when there are more than two inputs in order to overcome the technical complications that may arise when more than two inputs are included in production functions. As an additional suggestion for future research, the model proposed in this paper can also be applied to the approximation forms of the CES model such as Translog, which is an approximation of the CES function through a second-order Taylor expansion.

All in all, although production functions such as the CES are per se "technical" relationships between input quantities and output quantities, generalizing the CES to allow for the above-introduced components and real-world economic concepts while retaining its conventional desired properties will enrich the economic implications of this production function. This paper attempted to enhance the "economics" of the CES production function by adding these supplementary economic aspects to the standard form of the CES production function. We explored and proposed econometric possibilities for the CES production-function specification to allow for the flexibility to account for different types of input-driven output risks.

**Author Contributions:** Conceptualization, Ali Zeytoon-Nejad and Barry K. Goodwin; Methodology, Ali Zeytoon-Nejad and Barry K. Goodwin; Software, Ali Zeytoon-Nejad and Barry K. Goodwin; Validation, Ali Zeytoon-Nejad and Barry K. Goodwin; Formal analysis, Ali Zeytoon-Nejad and Barry K. Goodwin; Investigation, Ali Zeytoon-Nejad and Barry K. Goodwin; Data curation, Ali Zeytoon-Nejad and Barry K. Goodwin; Writing – original draft, Ali Zeytoon-Nejad and Barry K. Goodwin; Writing – review & editing, Ali Zeytoon-Nejad and Barry K. Goodwin; Visualization, Ali Zeytoon-Nejad and Barry K. Goodwin; Supervision, Barry K. Goodwin. All authors have read and agreed to the published version of the manuscript.

**Funding:** This research received no external funding.

**Data Availability Statement:** The data that support the findings of this study are available in Roger W. Hexem, Earl O. Heady, Metin Caglar (1974) A compendium of experimental data for corn, wheat, cotton and sugar beets grown at selected sites in the western United States and alternative production functions fitted to these data. Technical report: Center for Agricultural and Rural Development, Iowa State University. These data were derived from the following resources available in the public domain: https://babel.hathitrust.org/cgi/pt?id=wu.89031116783;view=1up;seq=3

**Conflicts of Interest:** The authors declare no conflict of interest.

**Appendix A. The Verification of the Main Properties of Just and Pope's Approach**

As argued by Just and Pope (1979), "the effects of input on output should not be tied to the effects of input on variability of output a priori". To achieve such a generality, they showed that one can work with a production function that in turn contains two separate functions, as shown in the following:

$$y = f(X) + h^{\frac{1}{2}}(X)\varepsilon, \quad E(\varepsilon) = 0, V(\varepsilon) = 1 \tag{A1}$$

Now, $E(y) = f(X)$ and $V(y) = h(X)$, and thereby, the effects of input on output are no longer tied to the effects of input on variability of output a priori. It can easily be affirmed that $\frac{\partial V(y)}{\partial x_i} = h_i(X)$, which indicates the fact that the sign of the expression $\frac{\partial V(y)}{\partial x_i}$ is no longer determined a priori, since it can take on any sign depending on the context of the problem. A similar verification, which has also been provided by Just and Pope (1979), can be provided for the marginal effect of input quantity on the variability of marginal productivity as follows:

$$-\frac{\partial y}{\partial x_i} = f_i(X) + \frac{1}{2}h^{-\frac{1}{2}}(X)h_i(X)\varepsilon \tag{A2}$$

$$V\left(\frac{\partial y}{\partial x_i}\right) = \frac{h_i^2(X)}{4h(X)} \tag{A3}$$

$$\frac{\partial V\left(\frac{\partial y}{\partial x_i}\right)}{\partial x_i} = \frac{h_i(X)[h(X)h_{ii}(X) - h_i^2(X)]^2}{2h^2(X)} \tag{A4}$$

Hence, the effects of input on variability of "output" and "marginal output" are not pre-determined anymore, since the signs of the above-mentioned expressions are no longer determined a priori. Interestingly, this result holds true even if $f(X)$ and $h(X)$ both follow the same production-function form, say, the CES production-function specification, which is the case of interest in the present paper. In this case, as an example, even if the estimated coefficient associated with the *i*th input turns out to be a negative parameter, then $h_i(X) < 0$, and as such, the case of a risk-reducing input is exemplified[19], which is not feasible to model with the original CES production-function specification.

It is also important to note that the original CES production function is indeed a special case of the Just and Pope variation of it (i.e., Equation (A1)). This becomes apparent if one takes into account the following possibility:

$$y = f^*(X)\varepsilon^* = f(X) + h^{\frac{1}{2}}(X)\varepsilon, \quad E(\varepsilon^*) = E(\varepsilon) = 0 \tag{A5}$$

where $f(X) \equiv h^{1/2}(X) \equiv f^*(X)E(e^{\varepsilon^*})$ and $\varepsilon \equiv e^{\varepsilon^*} - E(e^{\varepsilon^*})$. As such, the Just and Pope variant of the CES production function is simply more general than the original CES production-function specification.

**Appendix B. Details of the Estimation Procedure Associated with Just and Pope's Approach**

As discussed in Section 2, concerning the generalized variation of the CES production function that allows for the flexibility of input-driven output risk, we follow Just and Pope (1979)'s three-step approach. This appendix serves to describe and explain this approach in brief. For more information on this approach, you can see Just and Pope (1978, 1979). As argued by Just and Pope (1979) and discussed in Appendix A, to achieve such a generality, one can work with a production function that contains two separate functions, as shown in the following:

$$y = f(X) + h^{\frac{1}{2}}(X)\varepsilon, \ E(\varepsilon) = 0, V(\varepsilon) = 1 \tag{A6}$$

In this generalized production function, $E(y) = f(X)$ and $V(y) = h(X)$, and hence, the effects of input on output are no longer tied to the effects of input on variability of output a priori. For empirical purposes, suppose both $f(X)$ and $h(X)$ take the same functional form (of course, with different parameters); for the sake the present paper, say it is the original CES function. Therefore, the case of $f(X) = h^{1/2}(X)$ is a potential special case that recovers the original CES production function.

As the **first step** of this estimation procedure, the following regression model is estimated as a nonlinear, heteroscedastic regression of $y$ on $X$.

$$y_t = f(X_t, \alpha) + \varepsilon_t^*, \ E(\varepsilon_t^*) = 0, E(\varepsilon_t^* \varepsilon_\tau^*) = 0, for \ t \neq \tau \tag{A7}$$

where $\varepsilon_t^* = h^{1/2}(X)\varepsilon_t, E(\varepsilon_t) = 0, E(\varepsilon_t \varepsilon_\tau) = 0, for \ t \neq \tau$. Malinvaud (1970) has shown that when NLS is applied to the above regression model, it will result in consistent estimators of $\alpha$ and $f(X_t, \alpha)$.[20] This is in fact the end of the first step of the estimation procedure. If one is interested only in $f(X_t, \alpha)$ (i.e., the "mean" of output), they can stop here. However, there are multiple reasons for proceeding to the second and third steps as well. As outlined by Just and Pope (1979), the reasons could include the following: (1) learning more about the marginal effect of inputs usage on output risk, (2) performing more reliable hypothesis testing if there is a possibility of heteroscedasticity, and/or (3) gaining more efficiency in estimation, at least asymptotically.

As the **second step** of the estimation procedure, using $\hat{\alpha}$ from the first step (which is a consistent estimate of $\alpha$), $f(X_t, \alpha)$ can be consistently estimated by $f(X_t, \hat{\alpha})$. Therefore, $\varepsilon_t^*$ (which is equal to $h^{1/2}(X)\varepsilon_t$) can be estimated by the following:

$$\hat{\varepsilon}_t^* = y - f(X_t, \hat{\alpha}) \tag{A8}$$

Meanwhile, $E[(\varepsilon_t^*)^2] = E[h(X_t, \beta)\varepsilon_t^2] = h(X_t, \beta)$, which suggests $(\varepsilon_t^*)^2 = E[(\varepsilon_t^*)^2]u_t = h(X_t, \beta)u_t$ where $E[u_t] = 1$ by the definition of expectation.[21] That is to say, $\beta$ can be estimated through regressing $(\varepsilon_t^*)^2$ on $X_t$ in a nonlinear framework or a linear framework (since $\widehat{\varepsilon_t^*}$ consistently estimates $\varepsilon_t^*$).

As the **third step** of the estimation procedure, a weighted NLS regression of $y_t$ on $X_t$ in Equation (A7) with weights $h^{-1/2}(X_t, \hat{\beta})$ can achieve asymptotic efficiency in the estimation of $\alpha$. That is, an NLS estimate[22] of $\alpha$ can be found for the following model[23]:

$$y_t^* = f^*(X_t, \alpha) + \tilde{\varepsilon}_t \tag{A9}$$

where $y_t^* = y_t h^{-1/2}(X_t, \hat{\beta})$, and $f^*(X_t, \alpha) = f(X_t, \alpha)h^{-1/2}(X_t, \hat{\beta})$. Just and Pope (1978) have shown that such an estimator for $\alpha$ is consistent, asymptotically efficient, and unbiased under some conditions outlined in their paper.

In sum, Just and Pope (1979)'s approach can be summarized in the form of the following three-step estimation procedure:

(1) An NLS regression of $y_t$ on $f(X_t, \alpha)$, obtaining $\hat{\alpha}$.
(2) An NLS regression of $(\hat{\varepsilon}^*)^2 = (y_t - f(X_t, \hat{\alpha}))^2$ on $h(X_t, \beta)$, obtaining, $\hat{\beta}$.
(3) An NLS regression of $y_t^* = y_t h^{-1/2}(X_t, \hat{\beta})$ on $f^*(X_t, \alpha) = f(X_t, \alpha)h^{-1/2}(X_t, \hat{\beta})$, obtaining $\hat{\alpha}$.

**Appendix C. Details of the Estimation Procedure Associated with Just and Pope's Approach and Examining the Possibility of Heteroscedasticity in the Dataset**

As outlined by Just and Pope (1979) and discussed in this paper earlier, one may want to take the estimation of a production function to the second and third stages for three potential reasons, which include (1) learning more about the marginal effect of inputs use on output risk, (2) performing more reliable hypothesis testing if there is a possibility of heteroscedasticity, and/or (3) gaining more efficiency in estimation, at least asymptotically. Since the primary goal of the present paper is to explore the marginal effects of inputs use on output risk, we certainly have to proceed to the second and third stages of Just and Pope's estimation procedure; however, in this appendix, we examine the possibility of heteroscedasticity in the dataset to possibly find additional reasons to proceed to the second and third stages.

We start our preliminary investigations of the necessity of performing the second stage with a number of visual inspections and then proceed with more formal statistical examinations to detect any possible heteroscedasticity. In particular, we first examine whether or not there is any form of heteroscedasticity in the data, and if there is any visual indication of heteroscedasticity, we then examine more formally the existence of heteroscedasticity in the data by regressing the squared residuals *either* on the inputs *or* on the predicted output. Figure 1 depicts the squared residuals plotted against the observed independent variables for the estimated CES model of wheat at Yuma Mesa, AZ.

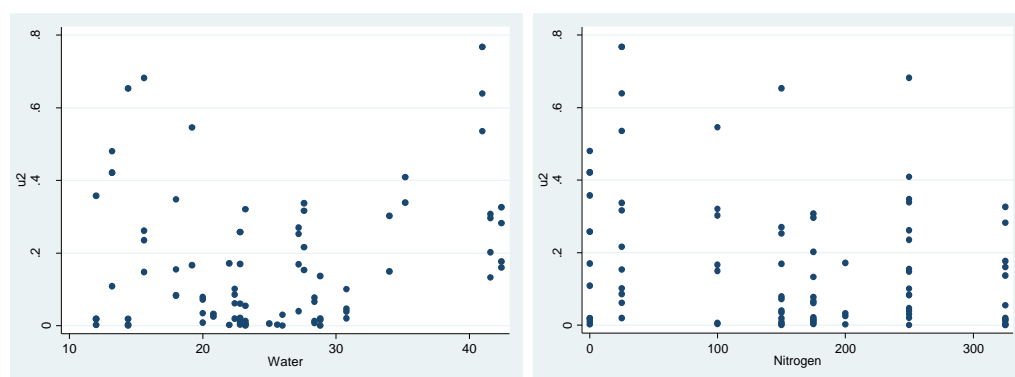

**Figure A1.** Squared Residuals Plotted against Independent Variables for the First-Stage Estimates of the Deterministic Component of Production Modeled by the CES Specification for Wheat at Yuma Mesa, AZ.

Figure 1 implies that there is some degree of heteroscedasticity in the data, meaning that the variance of the error term is not constant, and in fact, it varies with the usages of inputs. This can easily be seen if one considers an uneven envelope of residuals plotted against water (in which case it takes an *hourglass-shaped* heteroscedasticity form) and nitrogen (in which case it takes a roughly *decreasing* heteroscedasticity form, meaning that the variance of the squared residuals tends to decrease as the usage of nitrogen increases). As a result of these visual indications of heteroscedasticity, further statistical investigations of heteroscedasticity should be conducted. Table A1 provides the results of regressing the squared residuals on the independent variables (which is conceptually similar to the general idea of the Breusch–Pagan (BP) test for heteroscedasticity) for the estimated CES model of wheat production at Yuma Mesa, AZ, to see if there is any sort of linear relationship between the variance of the error term and the independent variables.

**Table A1.** Results of Regressing the Squared Residuals on the Independent Variables for the Estimated CES Model of Wheat Production at Yuma Mesa, AZ. (To detect any linear relationship between the variance of the error term and the independent variables in the first-stage estimates of the deterministic component of production).

| Source | SS | df | MS | | Number of obs | 88 |
|---|---|---|---|---|---|---|
| Model | 0.5661 | 2 | 0.2831 | | F(2, 85) | 9.1400 |
| Residual | 2.6335 | 85 | 0.0310 | | Prob > F | 0.0003 |
| | | | | | R-squared | 0.1769 |
| | | | | | Adj R-squared | 0.1576 |
| Total | 3.1996 | 87 | 0.0368 | | Root MSE | 0.1760 |
| $u^2$ | Coefficient | Standard Error | T | P > |t| | [95% Confidence | Interval] |
| Water | 0.0073 | 0.0023 | 3.26 | 0.0020 | 0.0029 | 0.0118 |
| Nitrogen | −0.0006 | 0.0002 | −3.50 | 0.0010 | −0.0010 | −0.0003 |
| Constant | 0.0861 | 0.0597 | 1.44 | 0.1530 | −0.0326 | 0.2047 |

As reported in Table A1, we can easily see that Prob > F = 0.0003, meaning that we are more than 99.9% confident that there is a statistically significant linear relationship between the variance of the error term and the independent variables. This is in fact an undesired observation, because it shows that the estimated standard errors and empirical significance levels (i.e., p-values) are unreliable and any statistical inference based on the results of the related regression model will be imprecise.

Another diagnostic plot that can be utilized to explore the existence of heteroscedasticity is the scatterplot of squared residuals versus the fitted values of the dependent variable. Figure A2 demonstrates this scatterplot for the estimated CES model of wheat at Yuma Mesa, AZ.

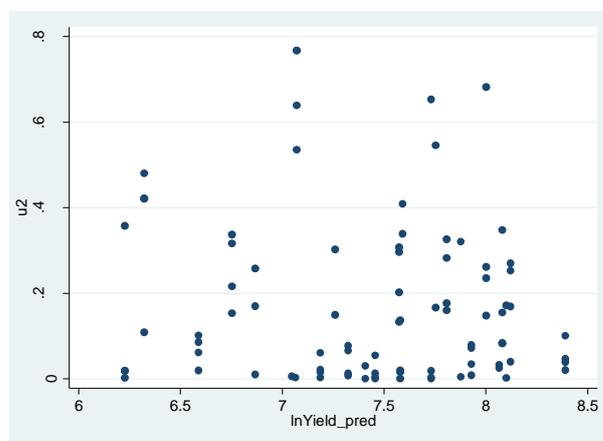

**Figure A2.** Residuals Plotted against Fitted Values for the First-Stage Estimates of the Deterministic Component of Production Modeled by the CES Specification for Wheat at Yuma Mesa, AZ.

Figure A2 implies that, when one considers the dependent variable to examine heteroscedasticity, taken together, there is a trivial degree of decreasing heteroscedasticity in the data, meaning that the variance of the error term is almost constant and does not change much with yield. This can be seen if one imagines the somewhat even envelope of residuals, whose width is almost constant for all values of yield. Despite this, a more formal statistical examination for heteroscedasticity could still be conducted for this case. Table A2 presents the results of regressing the residuals on the predicted dependent variable (i.e., fitted values) as well as the squares of the predicted dependent variable (which is conceptually similar to the general idea of the so-called White's General (WG) test[24] for heteroscedasticity) for the estimated CES model of wheat production at Yuma Mesa, AZ.

**Table A2.** Results of Regressing the Squared Residuals on the Predicted Dependent Variable and Square of It for the Estimated CES Model of Wheat Production at Yuma Mesa, AZ. (To detect any sort of relationship between the variance of the error term and the predicted dependent variable in the first-stage estimates of the deterministic component of production).

| Source | SS | df | MS | | Number of obs. | 88 |
|---|---|---|---|---|---|---|
| Model | 0.0521 | 2 | 0.0261 | | F(2, 85) | 0.7000 |
| Residual | 3.1475 | 85 | 0.0370 | | Prob > F | 0.4975 |
| | | | | | R-squared | 0.0163 |
| | | | | | Adj R-squared | −0.0069 |
| Total | 3.1996 | 87 | 0.0368 | | Root MSE | 0.1924 |
| $u^2$ | Coefficient | Standard Error | t | P > \|t\| | [95% Confidence | Interval] |
| Pred_lnYield | 0.3278 | 0.8323 | 0.39 | 0.6950 | −1.3271 | 1.9828 |
| Pred_lnYield$^2$ | −0.0251 | 0.0570 | −0.44 | 0.6610 | −0.1384 | 0.0882 |
| Constant | −0.8690 | 3.0233 | −6.88 | 0.7740 | −6.8802 | 5.1422 |

According to the results reported in Table A2, one cannot reject the hypothesis of no statistically significant relationship between the variance of the error term and the fitted values and the squared fitted values as a whole.[25] After all, although this second examination implies that the assumption of homoscedasticity cannot be rejected, we still take analysis to the second and third stages for two reasons: (1) There is an apparent inconsistency between the two sets of results reported in Tables A1 and A2, and more importantly, because (2) we are interested in observing the risk implications of the two inputs (i.e., irrigation water and nitrogen) separately, each of which could potentially have a different risk implication, which could help us better understand the workings and applicability of the proposed approach when more than one input is involved.

Figures A3 and A4 as well as Tables A3 and A4 report the same types of diagnostic plots and statistical examinations and in the same order as reported above, but for the case of Corn in Colby, KS.

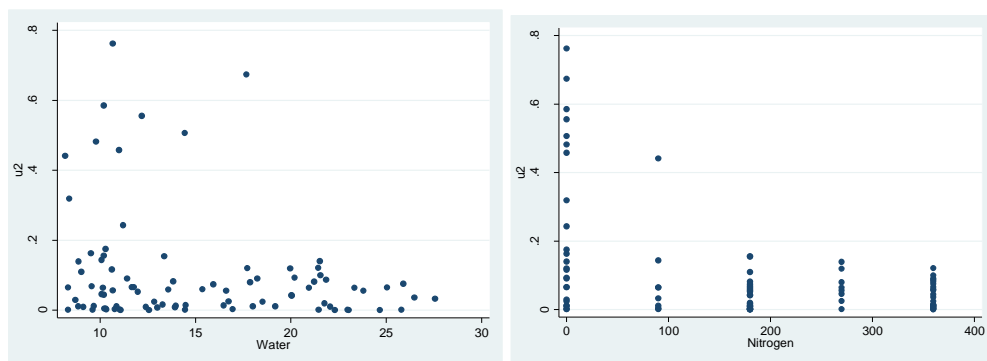

**Figure A3.** Squared Residuals Plotted against Independent Variables for the First-Stage Estimates of the Deterministic Component of Production Modeled by the CES Specification for Corn in Colby, KS.

Figure A3 shows that there is a noticeable degree of *decreasing* heteroscedasticity, implying that the variance of the squared residuals tends to decrease as the usages of inputs increase. As a result of these visual indications of heteroscedasticity, further statistical investigations of heteroscedasticity are undertaken below. Table A3 provides the results of regressing the squared residuals on the independent variables for the estimated CES model of corn production in Colby, KS to find out if there is any sort of linear relationship between the variance of the error term and independent variables.

**Table A3.** Results of Regressing the Squared Residuals on the Independent Variables for the Estimated CES Production Specification for Corn in Colby, KS. (To detect any linear relationship between the variance of the error term and the independent variables in the first-stage estimates of the deterministic component of production).

| Source | SS | df | MS | | Number of obs. | 88 |
|---|---|---|---|---|---|---|
| Model | 0.5033 | 2 | 0.2516 | | F(2, 85) | 12.5600 |
| Residual | 1.7033 | 85 | 0.0200 | | Prob > F | 0.0000 |
| | | | | | R-squared | 0.2281 |
| | | | | | Adj R-squared | 0.2099 |
| Total | 2.2066 | 87 | 0.0254 | | Root MSE | 0.1416 |
| $u^2$ | Coefficient | Standard Error | t | P>|t| | [95% Confidence | Interval] |
| Water | −0.0043 | 0.0028 | −1.51 | 0.1340 | −0.0099 | 0.0013 |
| Nitrogen | −0.0005 | 0.0001 | −4.46 | 0.0000 | −0.0007 | −0.0003 |
| Constant | 0.2547 | 0.0464 | 5.49 | 0.0000 | 0.1624 | 0.3470 |

The results in Table A3 indicate that there is a statistically significant linear relationship between the variance of the error term and the independent variables. Hence, the estimated standard errors are unreliable, and any statistical inference based on the results of the related regression model will be imprecise.

As an additional diagnostic plot to explore the existence of heteroscedasticity, one can look at the scatterplot of the squared residuals versus the fitted values of the dependent variable. Figure A4 demonstrates such a scatterplot for the estimated CES model of corn in Colby, KS.

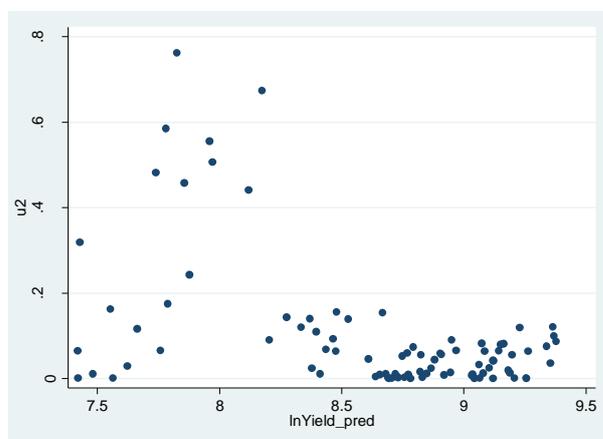

**Figure A4.** Residuals Plotted against Fitted Values for the First-Stage Estimates of the Deterministic. Component of Production Modeled by the CES Specification for Corn in Colby, KS.

Figure A4 suggests that there is a considerable degree of *decreasing* heteroscedasticity in the data, meaning that the variance of yield is larger for smaller values of yield than that for greater values of it. A more formal statistical examination for heteroscedasticity is performed below. Table A4 provides the results of regressing the residuals on the fitted values as well as the squared fitted values for the estimated CES model of corn production in Colby, KS.

**Table A4.** Results of Regressing the Squared Residuals on the Predicted Dependent Variable and Square of It for the Estimated CES Model of Corn Production in Colby, KS. (To detect any sort of relationship between the variance of the error term and the predicted dependent variable in the first-stage estimates of the deterministic component of production).

| Source | SS | df | MS | | Number of obs. | 88 |
|---|---|---|---|---|---|---|
| Model | 0.5185 | 2 | 0.2592 | | F(2, 85) | 13.0500 |
| Residual | 1.6881 | 85 | 0.0199 | | Prob > F | 0.0000 |
| | | | | | R-squared | 0.2350 |
| | | | | | Adj R-squared | 0.2170 |
| Total | 2.2066 | 87 | 0.0254 | | Root MSE | 0.1409 |
| $u^2$ | Coefficient | Standard Error | t | P > \|t\| | [95% Confidence | Interval] |
| Pred_lnYield | 0.1736 | 0.8765 | 0.20 | 0.8430 | −1.5692 | 1.9164 |
| Pred_lnYield$^2$ | −0.0186 | 0.0520 | −0.36 | 0.7220 | −0.1219 | 0.0848 |
| Constant | −0.0079 | 3.6808 | 0.00 | 0.9980 | −7.3263 | 7.3105 |

The results reported in Table A4 indicate that there is a statistically significant relationship between the variance of the error term and the dependent variable for the case of modeling corn production in Colby, KS. Again, this is indeed an undesired observation, since any statistical inference based on the results of the related regression model will be imprecise and the associated standard errors and empirical significance levels (i.e., p-values) are unreliable.

The bottom-line conclusion that can be drawn from all of the preceding diagnostic examinations is that (1) there exists a sizable degree of heteroscedasticity in the data, implying that any statistical inference based on the results of the related regression model could be imprecise, and more importantly, (2) the type of the existing heteroscedasticity in most cases is *decreasing* in inputs, which cannot be addressed and accounted for accurately when one uses the original CES production-function specification, as shown and proven in Section 2 by Equation (3).

**Notes**

1. Other types of risks associated with the process of production can include the market prices of inputs, the market price of output, wildfire- and weather-related risks, pests and disease, infrastructure malfunction, risks associated with production-related regulations and policy shocks, as well as technical and technological risks.
2. As Just and Pope (1979) elaborate, in such circumstances and under risk aversion, "the true utility loss associated with higher risk (at the lower input level) will be greater than when the risk effect is incorrectly estimated as a reduction in variability".
3. As Just and Pope (1979) put it, "Consider, for example, overcapitalization in grain harvesting. The use of large (and fast) harvesting equipment, as opposed to smaller (and slower) equipment, usually leads to less variability of output (because of random weather conditions which can destroy a ripe crop before harvest".
4. To see some alternatives to this Just and Pope specification, you can see Just and Pope (1978), which is a comprehensive paper discussing various aspects of and introducing different alternatives to this particular specification.
5. To verify that this specification has the property that the signs of neither Equation (3) nor Equation (6) are determined a priori anymore, you can see Appendix A, which verifies this property in great detail.
6. One may want to estimate a generalized variant of the CES production-function specification that incorporates simultaneously both the generalization proposed in this paper and that introduced in Zeytoon-Nejad et al. (2022), which incorporates input thresholds within the structure of the CES production function. Zeytoon-Nejad et al. (2022) provide an appendix detailing the technical aspects of such a joint estimation. They also provide a comprehensive literature review on the CES production function, as well as the evolution of its different variants and generalizations.
7. Henningsen and Henningsen (2011) provide a longer list of estimation methods that can be applied for the estimation of the CES function. However, since the other methods have been of less popularity among empirical economists, we do not report them in this paper. For more information on these methods, you can see Henningsen and Henningsen (2011).
8. The NLS approach routinely works well in many empirical applications with real-world data. However, under some circumstances, it does not perform well, primarily due to three potential reasons: (1) non-convergence even after numerous iterations due to lack of a unique minimum of the sum-of-squares, (2) convergence to a local minimum due to improper choices of initial values, or (3) generating theoretically nonsensical parameter estimates. In such cases, one should resort to alternative ways of estimating the CES production function, a list of which is provided in Henningsen and Henningsen (2011).
9. An alternative to time-dummy variables for the purpose of investigating the effects of weather-related variables on all contemporaneous observations is to use a variance components procedure. For further discussion of variance components procedures, you can see Wallace and Hussain (1969), Maddala (1971), and Just and Pope (1979).
10. That is, logarithms are used for the deterministic part separately from those for the stochastic part. This is because the estimations of the two function components are, in practice, performed in a sequence of stages and separately from one another. In other words, in none of the stages are both components of the function estimated, and instead, in each stage, only one component of the function is estimated, as outlined by Just and Pope (1979)'s three-step estimation procedure.
11. As a result, we need to disjoin the effect of marginal use of input on the *mean* of output from that on the *variability* of output, as suggested by Just and Pope (1979). Although they have applied their approach for the case of Cobb–Douglas and Translog production-function specifications, as they point out in their seminal papers published in 1978 and 1979, the same general approach can be applied to other production-function specifications such as the CES specification as well.
12. For the second stage, functional forms other than the one employed for the first stage can also be utilized, as explained by Just and Pope (1978). As a result, for the purpose of the second stage, we could have used functional forms other than the CES specification. However, we still preferred to use the CES specification for the sake of consistency, its flexibility, and more importantly, because of the fact that the focus of the present paper is on the exposition of the many capabilities of the CES specification. After all, as reasonably pointed out by Griffin et al. (1987), in many production analyses, selecting functional forms in a totally objective manner is almost impossible, and formalization of the selection process mostly requires deliberate choice and frank presentation.
13. Additionally, and on a related note, the fact that the estimated constant terms (B's) are negative is not troublesome, either. This is because these coefficients are no longer interpreted as TFP. Instead, they can be interpreted as the portion of output "variance" that is not explained by the amount of inputs used in production.
14. These scatterplots differ, in essence, from those provided in Appendix C, in that these ones are drawn using the actual data and thus represent the heteroscedasticity existing purely in the data. In fact, in the scatterplots presented in Appendix C, the change in variance could be due to a potential heteroscedasticity inherent in the data and/or a potential misspecification. It is known that model misspecifications such as wrong functional forms or omitted variables can also produce heteroscedasticity. Under such circumstances, if the model is specified correctly, the patterns of heteroscedasticity may disappear. Thus, in the scatterplots presented in Appendix C, the change in variance could be due to a potential heteroscedasticity inherent in the data and/or a potential misspecification, and as such, those scatterplots are less relevant to our topic of interest, i.e., input-driven output risk, which is defined as the marginal effect of input use on the variance of output. In contrast, the scatterplots presented here are more relevant to the study of input-driven output risk, which can directly be observed in these scatterplots regardless of and separately from a potential problem of misspecification.
15. More precisely, it should be noted that these scatterplots demonstrate the marginal effects of inputs use on output *variance* simultaneously with the marginal effects of inputs use on output *mean*, which prevents one from purely observing the former

  effect (i.e., the effect of inputs use on output *variance*, which is the main purpose of the second stage of Just and Pope's approach). As a result, the desired comparisons cannot be absolutely effectively made using these visual checks, since imaging the mean and its trend while considering variance and its trend at the same time is a hard task to carry out. Despite this, since we find the provision of these visual checks still valuable, we have included and discussed these comparisons here.

16. The nearly zero marginal effect of nitrogen on the variability of output in this dataset is somewhat noticeable if one looks at the initially increasing variances of data points in the lower levels of nitrogen and decreasing variances of yield with higher levels of nitrogen. In fact, these two effects somehow offset each other, and the estimated (net) effect of nitrogen use on the variability of output becomes close to zero.
17. The increasing effect of water on the variability of output in this dataset is somewhat noticeable if one takes into account the clustered data points in the low levels of water and the relatively more dispersed data points associated with higher levels of water irrigation.
18. The nearly zero marginal effect of nitrogen on the variability of output in this dataset is somewhat noticeable if one looks at the initially decreasing variances of data points in the lower levels of nitrogen and increasing variances of yield with higher levels of nitrogen. In fact, these two effects somehow offset each other, and the estimated (net) effect of nitrogen use on the variability of output becomes close to zero.
19. In this paper, the case of risk-reducing input turned out to be true for the case of nitrogen in our empirical applications of wheat at Yuma Mesa, AZ, and corn in Colby, KS, although the negative estimated coefficients are close to zero. In Just and Pope (1979), this result turned out to be true for the case of fertilizer in their empirical application of oats using the well-known Day dataset.
20. For more details on this, see Just and Pope (1978).
21. To understand this better, see Hildreth and Houck (1968) and Theil (1971).
22. Although the NLS method is a powerful tool for fitting nonlinear models such as the CES production function, it has some limitations and weaknesses, too. Zeytoon-Nejad et al. (2022) have discussed such potential limitations of the NLS method as it may relate to the estimation of the CES production function. It is also important to note that, in general, the choice of the starting values needed for the optimization algorithms of the NLS method may cause the software to end up with incorrect estimates due to a convergence to local optima or a possibility of convexity of the objective function of the NLS optimization problem. As diagnostics tests to rule out this possibility in this study, a very wide range of values (in the domain of theoretically meaningful ranges of starting values) were tried as the starting value of the NLS optimization algorithms, and the change in that choice did not change any of the results, indicating that the estimates are the result of the optimization algorithm when meeting global optima and not local ones.
23. For an extensive discussion of the identification of the CES production function as a nonlinear regression model, and also to see the mathematical proof of the global identifiability of the Generalized CES (GCES) production function, you can see Zeytoon-Nejad et al. (2022), in which both of these matters have been comprehensively discussed in the form of two extensive appendices.
24. The WG test for heteroscedasticity is a flexible test and identifies almost any pattern of heteroscedasticity. It even allows the independent variable to have a nonlinear effect on the error variance, while the BP test assumes that heteroscedasticity is a linear function of independent variables.
25. This result is somewhat different from when we considered heteroscedasticity by examining the relationship between the variance of the error term and the individual inputs. The two scatterplots presented in Figure A1 suggest that one possible reason for this apparent inconsistency could be that the two different heteroscedasticity situations existing in the relationship between the residuals and inputs somehow offset each other, so when one looks at the heteroscedasticity existing in the relationship between the error term and the dependent variable, taken together, heteroscedasticity is no longer observed, and overall, homoscedasticity in this case cannot be rejected.